\documentclass[onecolumn,amsmath,amssymb,nofootinbib,12pt]{article}
\pdfoutput=1
\usepackage{jheppub}

\usepackage{graphicx}
\usepackage{dcolumn}
\usepackage{bm,psfrag}

\usepackage{subfigure}
\usepackage{float}
\usepackage{tensor}

\newcommand{\ben}{\begin{equation}}
\newcommand{\een}{\end{equation}}
\newcommand{\be}{\begin{equation}}
\newcommand{\ee}{\end{equation}}
\newcommand{\bea}{\begin{eqnarray}}
\newcommand{\eea}{\end{eqnarray}}
\newcommand{\ba}{\begin{eqnarray}}
\newcommand{\ea}{\end{eqnarray}}
\newcommand{\nn}{\nonumber}

\newcommand{\beq}{\begin{equation}}
\newcommand{\eeq}{\end{equation}}
\newcommand{\beqa}{\begin{eqnarray}}
\newcommand{\eeqa}{\end{eqnarray}}
\newcommand{\beqar}{\begin{eqnarray*}}
\newcommand{\eeqar}{\end{eqnarray*}}

\newcommand{\cO}{{\cal O}}

\def\t6 {T_\mt{D6}}

\newcommand{\mt}[1]{\textrm{\tiny #1}}

\newcommand{\cA}{{\cal A}}

\newcommand{\vk}{{\vec{k}}}
\newcommand{\vx}{{\vec{x}}}

\def\cale         {{\cal E}}

\def\calo         {{\cal O}}

\def\ee           {{\rm e}}

\def\sqr#1#2{{\vcenter{\vbox{\hrule height.#2pt
 \hbox{\vrule width.#2pt height#1pt \kern#1pt
 \vrule width.#2pt}\hrule height.#2pt}}}}

\def\ee{\cale}

\def\aa1{\phi}
\def\cc1{\psi}

\def\nn{\nabla_\nu}

\newcommand{\dt}{\delta t}

\def\cO{{\cal{O}}}
\def\cE{{\cal{E}}}
\def\vx{{\vec{x}}}
\def\vk{{\vec{k}}}
\def\cA{{\cal{A}}}
\def\cJ{{\cal{J}}}
\def\cP{{\cal{P}}}
\def\cD{{\cal{D}}}

\def\cC{{\cal{C}}}
\def\dt{\delta t}
\def\Psis{\Psi_s}
\def\calo{{\cal{O}}}

\begin{document}


\title{Old and New Scaling Laws in Quantum Quench \footnote{Based on talk at Nambu Symposium, University of Chiacgo, March 2016.}}

\author{Sumit R. Das}
\affiliation{Department of Physics and Astronomy, University of Kentucky,\\ 
\vphantom{k}\ \ Lexington, KY 40506, USA}

\emailAdd{das@pa.uky.edu}

\date{\today}

\abstract{The response of a many body system to a time dependent coupling which passes through or approaches a critical point displays universal scaling behavior. In some regimes, scaling laws have been known since the 1970's. Recently holographic techniques have been used to understand the origins of such scaling. Along the way, new scaling behaviors in other regimes have been found in holographic models, which have later been shown to hold in a generic field theory regardless of holography. This contribution summarizes recent work on these various scaling regimes. }

\date{\today}

\maketitle


\section{Introduction}

It is a great honor for me to speak in this symposium in memory of my mentor Yoichiro Nambu. I have been truly fortunate to be his student and get some glimpses of the depth of his thinking and of course his profound originality. The four years I spent in Chicago led to a lifelong relationship during which he continued to shape my taste in physics. 

Nambu liked problems which transcend traditional boundaries in physics. Quantum quench is one such problem which has applications to a wide range of areas of physics - from cosmology to the physics of quark-gluon liquid to the physics of cold atom systems. 

The statement of the problem is quite simple. Consider a system whose hamiltonian contains a time dependent parameter $\lambda (t)$ which asymptotes to constant values at early and late times and changes over a time scale $\delta t$. Following standard terminology  I will call this quantum (thermal) quench when the initial state is the vacuum (thermal state), {\em regardless} of the rate of change.  
The time dependent coupling $\lambda (t)$ excites the system, and the question is to determine the characteristics of physical quantities after the quench is over. In the following I will deal exclusively with {\em global quench}, i.e. space translation is not broken. 

One motivation behind studying such a problem is to understand thermalization. Starting from a ground state does the late time state resemble a thermal ensemble ? If so, in what sense ? In recent years experiments with cold atom systems and heavy ion collisions are beginning to probe this process which lies at the heart of statistical mechanics. A second interest lies in studying properties of cosmological fluctuations. Here the expanding universe render couplings effectively time dependent and the problem of quench becomes the problem of particle production in time dependent geometries. In this talk I will concentrate of a third interest : critical dynamics. When the quench process passes through or approaches a critical point, one would expect that the subsequent time evolution of the system will reflect universal properties of the critical point. Our aim is to explore such universal behavior and understand their origins.

An early example of such universal behavior is Kibble-Zurek scaling. This was discovered while trying to understand defect formation during thermal phase transitions in cosmology \cite{kibble}, extended to condensed matter systems driven through a critical point by a time dependent source some time later \cite{zurek}, and to quantum critical transitions more recently \cite{more}.
Making what may appear to be rather drastic assumptions, Kibble and Zurek showed that the density of defects  scales in a simple way with the rate of change of the parameters of the theory. The exponents are determined by the equilibrium critical exponents. 
It turns out that defect density is just one of the quantities which display scaling behavior. We will call this kind of scaling {\em Kibble-Zurek scaling}, regardless of whether this results from the specific mechanism proposed by \cite{kibble,zurek}.

Kibble-Zurek scaling appears for {\em slow quenches}. In quantum quench, this means that we start from a gapped phase with a rate of change of the coupling slow compared to the initial gap. Subsequently the coupling approaches a value which would correspond to a critical point in equilibrium. Clearly the initial time evolution will be adiabatic. However when the coupling approaches the critical point, the instantaneous gap becomes small and adiabaticity inevitably breaks down. Let us assume that this happens close enough to the critical point where the coupling changes with time in a power law fashion
\ben
\lambda(t) - \lambda_c \sim \lambda_0 \left( \frac{t}{\delta t} \right)^{r}
\label{1-1}
\een
where $\lambda_c$ is the value of the coupling where the equilibrium system is critical, and $r$ is some integer. In this regime the instantaneous energy gap $E_g (t)$ scales with equilibrium critical exponents
\ben
E_g (t) \sim |\lambda(t) - \lambda_c|^{z\nu}
\label{1-2}
\een
where $z$ is the dynamical critical exponent and $\nu$ is the correlation length exponent. Then adiabatic evolution requires
\ben
\frac{1}{[E_g (t)]^2}\frac{d E_g (t)}{dt} \ll 1
\label{1-3}
\een
Adiabaticity will break when this quantity becomes of order one. This happens at a time $t = t_{KZ}$ (called the Kibble-Zurek time) which follows from (\ref{1-1})-(\ref{1-3})
\ben
t_{KZ} \sim (\delta t)^{\frac{z\nu r}{z \nu r +1}}
\label{1-4}
\een
The instantaneous correlation length at this time is
\ben
\xi_{KZ} = \xi (t_{KZ}) \sim t_{KZ}^{1/z}
\label{1-5}
\een
For $t > t_{KZ}$ it is generally difficult to follow the time
evolution. Kibble (and Zurek) proposed that one may proceed by
assuming that soon after $t = t_{KZ}$ the system becomes {\em diabatic}, i.e.
all physical quantities remain exactly the
same as it was at $t= t_{KZ}$. It then resumes adiabatic evolution
once the system comes out of the critical region \footnote{Recently it has been found that this is not quite true : there is a period of non-adiabatic coarsening after the end of the quench \cite{chesler-liu}}. If we assume, in
addition, that the only length scale in the critical  region is $\xi_{KZ}$
,expectation values of operators would scale as powers
of  $\xi_{KZ}$ \cite{more}
\ben
<\cO> \sim \xi_{KZ}^{-\Delta}
\label{1-6}
\een
where $\Delta$ is the conformal dimension of the operator $\cO$.

A slightly improved version of the scaling hypothesis states that expectation values and correlation functions do not quite stay frozen in this regime. Rather they are governed by {\em scaling functions} \cite{qcritkz}
\bea
<\cO (t)> & \sim & \xi_{KZ}^{-\Delta}~F_1(t/t_{KZ}) \nonumber \\
<\cO (\vx,t) \cO (0,t^\prime)> & \sim & \xi_{KZ}^{-2\Delta}~F_2(|\vx|/\xi_{KZ}, t/t_{KZ}, t^\prime/t_{KZ})
\label{1-7}
\eea
where $F_1$ and $F_2$ are the scaling functions.

The assumptions which go into the general derivation of KZ scaling appear drastic. Nevertheless, such scaling has been shown to hold in specific solvable models (where the scaling relations follow without these assumptions) and there are indications that such scaling has indeed been observed in some experiments. Unfortunately, there is no general theoretical framework which explains why these assumptions are valid. As will be discussed below, mapping the problem to classical gravity in one higher dimension using the holographic correspondence has led to some insight \cite{holo-kz1} -\cite{chesler-liu}.

Universal scaling laws have also been found in the opposite regime of
an {\em abrupt} or {\em instantaneous} quench. In this case, the coupling changes suddenly from one value to another at some time e.g. at $t=0$. During this process the state does not change. However if the initial state is the ground state of the initial hamiltonian $H_0$, it is an excited state of the new hamiltonian $H_1$ and evolves non-trivially according to $H_1$. Suppose $H_0$ is gapped, and $H_1$ is critical. Calabrese and Cardy \cite{cc2,cc3} argued that in this case, for purposes of IR quantities the state $|\psi_0>$ at the time of the quench may be approximated by
\ben
|\psi_0> \approx {\rm exp}[-\tau_0 H_1] |B>
\label{1-8a}
\een
where $\tau_0$ is roughly the inverse of the mass gap of $H_0$ and $|B>$ is a boundary state in the conformal field theory described by $H_1$. In $1+1$ dimensions, powerful methods of boundary conformal field theory can be then used to obtain universal properties of correlation functions. For example the one point function of a primary field with conformal dimension $\Delta$ behaves as
\ben
<\cO (t)> \sim {\rm exp}[ - \frac{\pi \Delta}{2\tau_0} t]
\label{1-8}
\een
Thus the ratio of the relaxation times of two different operators is purely given by the inverse of the ratio of their conformal dimensions, which is universal. Another example is the entanglement entropy $S_{EE}(t,L)$ of an interval of size $L$. This behaves as
\ben
S_{EE}(t,L) = \frac{\pi c t }{6\tau_0} \theta (t - L/2) + \frac{\pi c L}{12 \tau_0} \theta (L/2-t)
\label{1-9}
\een

Inbetween these two extreme limits lies a regime of quench rate which is {\em slow compared to the UV cutoff scale, but fast compared to physical mass scales.} We will call this "fast quench". This regime was first studied using holographic methods in \cite{numer} and \cite{fastQ}. These methods were used to study boundary relativistic actions of the form
\ben
S = S_{critical} - \int dt \int d^{d-1}x~\lambda(t) \cO(\vx,t)
\label{1-10}
\een
where $\cO$ is a relevant operator with conformal dimension $\Delta$. It was found that the {\em renormalized} $<\cO>$ and energy density $<\cE>$ soon after the quench scale in a universal fashion
\ben
<\cO> \sim (\delta t)^{d-2\Delta}~~~~~~~~<\cE> \sim (\delta t)^{d-2\Delta}
\label{1-11}
\een
with logarithmic enhancement in even dimensions. This is a new kind of universal scaling behavior.

As we will explain below, it turns out that this universal behavior is completely general for {\em any} field theory, and is not restricted to those field theories which have gravity duals \cite{dgm1} -\cite{dgm4}.

In the following I will summarize some salient aspects of these various scaling regimes and the cross-over between them.

\section{The Holographic Set-up}

Under suitable conditions, the AdS/CFT correspondence relates strongly coupled field theories in $d$ space-time dimensions with gravitational theories in $d+1$ dimensional AdS space-time with appropriate bulk fields turned on.( It is useful to think of the field theory as a deformation of a conformal field theory) . The field theory lives on the boundary of this AdS space.  

We will consider regimes where the bulk theory is weakly coupled.  Consider the Poincare patch of AdS with metric (The AdS scale has been set to unity)
\ben
ds^2 = \frac{1}{z^2} [ -dt^2 + dz^2+d\vx^2]
\label{2-1}
\een
Then the boundary is at $z=0$. In this case, for each operator in the conformal field theory there is a field in the $d+1$ dimensional bulk, e.g.

\begin{table}[!h]
\centering
\begin{tabular}{|c||c||c|}
\hline
{\rm Scalar} & $\cO (\vx,t)  \leftrightarrow \phi(z,\vx,t) $ & {\rm scalar field}\\
\hline
{\rm vector current} & $J_\mu (\vx,t)  \leftrightarrow A_\mu(z,\vx,t) $ & {\rm gauge field}\\
\hline
{\rm Energy-Momentum Tensor} & $ T_{\mu\nu}(\vx,t)  \leftrightarrow h_{\mu\nu}(z,\vx,t) $ & {\rm metric perturbations}\\
\hline
\end{tabular}
\end{table}
When the metric in the bulk is exactly $AdS$ and there are no other bulk fields turned on, the dual field theory is a conformal field theory in its vacuum state. When bulk fields are turned on, their
asympototic behavior near the boundary determines the nature of deformation of the CFT (i.e. the sources) as well as the nature of the state of the theory.
For example a scalar field with mass $m$ in the bulk has an expansion near the boundary of the form
\ben
\phi(\vx,t,z) \sim z^{d-\Delta}[\lambda(\vx,t) + O(z^2)] + z^\Delta[ A(\vx,t) + O(z^2)]
\label{2-2}
\een
the dual field theory has an action
\ben
S = S_{CFT} - \int d^{d-1}x dt~\lambda (\vx,t) \cO( \vx,t)
\label{2-3}
\een
where the operator $\cO$ has a conformal dimension $\Delta$, while its expectation value is given by
\ben
<\cO (\vx,t) > = A(\vx,t)
\label{2-4}
\een
The dimension $\Delta$ is related to the mass of the scalar by the standard relation
\ben
\Delta= \frac{d}{2} + \sqrt{\frac{d^2}{4} +m^2}
\label{2-4a}
\een

In an analogous fashion the asymptotic behavior of a bulk Maxwell field,
\ben
A_\mu \sim \cA_\mu (\vx,t)  (1+ O(z^2)) + z^{d-2} \cJ_\mu (\vx,t) (1+ O(z^2))
\label{2-5}
\een
signifies that the CFT is deformed by a term $\int d^{d}x \cA_\mu J^\mu (\vx,t)$ while the expectation value of the current operator is given by $\cJ_\mu (\vx,t)$. In particular a nonzero constant $\cA_0 = \mu$ is a chemical potential for the (global) charge $J^0$ while $\cJ_0$ is the charge density.

A nonzero expectation value in the absence of a source may signify an excited state. A nontrivial state of particular interest is a thermal state. The dual of this is a black hole with a Hawking temperature $T$.
Alternatively - as we will see - this could happen due to 
spontaneous symmetry breaking.  

\section{Holographic Quantum Quench and Kibble-Zurek}

It is clear that the gravity dual description of quantum quench in the field theory on the boundary is time evolution caused by a time dependent boundary condition. Early work along these lines used time dependent boundary conditions to construct toy models  of cosmological singularities \cite{cosmo,cosmo2}. The idea is to have a boundary theory coupling start from large constant value, dip down to a very small value and rise up again to some other constant value. The regime of small coupling would then correspond to a spacelike or null region in the bulk of high curvature - which appear as bulk singularities \footnote{For related approaches to holographic cosmology, see \cite{cosmo3}.}. While time evolution cannot be computed in the bulk, the boundary theory may admit a well defined time evolution . While models with null singularities can be found where the boundary theory evolution is well defined, there is no clear conclusion for examples leading to spacelike singularities, though there has been some recent interesting results in such models \cite{recentcosmo}.

If the boundary field theory starts out in its vacuum state with a zero source and a time dependent coupling is turned on at some finite time, and subsequently turned off at some later time, the resulting injection of energy produces a disturbance which propagates into the bulk. Typically this leads to formation of a black hole horizon. In the boundary field theory this manifests itself as thermalization, with a temperature which is equal to the Hawking temperature of the black hole. This process has been studied extensively over several years \cite{bh1}-\cite{bh6a} in various contexts - particularly as models of thermalization in heavy ion collisions. What is important is the formation of an apparent horizon : in fact apparent horizons formed due to such quench processes in probe brane models in AdS/CFT lead to thermality in flavor sectors \cite{bh5}. An important aspect in this context is the investigation of the dynamics of holographic entanglement entropy \cite{hent}. The gravity dual description of abrupt quench states like (\ref{1-8a}) has been studied in \cite{hartmanmaldacena}. Black hole formation in the bulk and resulting Vaidya type metrics have been directly shown from the dual conformal field theory in \cite{hartman1}.

Our interest here is in quantum quenches which involve critical points. This means we need to consider a gravity dual description of a model which has a quantum critical point and dynamically go across this by a suitable time dependent boundary condition. This program was carried out in \cite{holo-kz1} -\cite{holo-kz-other}. In the following, I will discuss the model used in \cite{holo-kz2} and \cite{holo-kz4} and summarize the main results.

\subsection{A holographic superfluid}

The model we will discuss is a modification of the bottom-up model of \cite{holosuper}. The bulk action in five space-time dimensions is given by
\ben
S = \int d^{5}x {\sqrt{g}} \left[ \frac{1}{2\kappa^2} \left( R +
  \frac{d(d+1)}{L^2}\right) - \frac{1}{4}F_{\mu\nu}F^{\mu\nu} -
  \frac{1}{\lambda} \left( |\nabla_\mu \Phi - iqA_\mu \Phi|^2 - m^2
  |\Phi|^2 - \frac{1}{2} |\Phi|^4 \right) \right] \ ,
\label{3-1}
\een
where $\Phi$ is a complex scalar field and $A_\mu$ is an abelian gauge
field, and the other notations are standard. Henceforth we will use $L=1$
units. One of the spatial
directions, which we will denote by $\theta$ is
compact. In \cite{holosuper} there was no self-interaction of the scalar. The presence of a self interaction allows an extreme probe approximation
\ben
\lambda \gg q^2 \ ,~~~~~~\lambda \gg \kappa^2 \ .
\label{probe}
\een
where the scalar field is a probe field, whose back-reaction to both the metric and the gauge field can be ignored. We will first consider this approximation and then incorporate the back-reaction. This is inspired by the setup of
\cite{liu1}.

Let us first consider the background with $\Phi = 0$. It is well known that there are two possibilties. At zero temperature, the first is the $AdS_5$ soliton
\bea
ds^2 & = & \frac{dr^2}{r^2 f_{sl}(r)} + r^2 \left( -dt^2 +
\sum_{i=1}^{2} dx_i^2 \right) + r^2 f_{sl}(r) d\theta^2 \ ,\nn \\
f_{sl}(r) & = & 1 - \left( \frac{r_0}{r} \right)^{d+1}, ~~~~~~
A_t  =  \mu 
\label{3-2}
\eea
with constant parameters $\mu$ and $r_0$. The periodicity of $\theta$ in
this solution is
\ben
\theta \sim \theta + \frac{\pi}{r_0} \ ,
\label{3-3}
\een
 The second solution is an extremal
$AdS_{5}$ charged black brane
\bea
ds^2 & = & -r^2 f_{bh}(r) dt^2 + \frac{dr^2}{r^2 f_{bh}(r)}+r^2 \left(
\sum_{i=1}^{2} dx_i^2 + d\theta^2 \right) \ , \nn \\
f_{bh}(r) & = & 1 - \frac{\mu^4}{12 r^4} + \frac{\mu^6}{108 r^6},
~~~~A_t  = \mu \left[ 1 - \frac{\mu^2}{6 r^2} \right] \ .
\label{3-4}
\eea
The period of $\theta$ is arbitrary. As shown in
\cite{holosuper}, this system undergoes a phase transition
between these two solutions when 
\ben
\mu = \mu_{c2} = \sqrt{2}(3)^{1/4} r_0
\label{3-5}
\een
The AdS soliton is stable when $\mu < \mu_{c2}$. These two solutions and the first order transition between them extends to finite temperatures. They also extend to arbitrary dimensions $d > 1$.

We now show that there is a value of $\mu = \mu_{c1}$ inside the soliton phase such that for $\mu_{c1} < \mu < \mu_{c2}$ this $\Phi = 0$ solution is not thermodynamically stable for a range of masses $-4 \leq m^2 \leq -3$. Note that this is the window of masses where there are two possible quantizations. 

In the following we will rescale all dimensionful quantities appropriately and set $r_0 = 1$. In the extreme probe approximation (\ref{probe}) we need to consider only the scalar wave equation in the soliton background. For fields which depend only on $t$ and $r$, the equation of motion is
given by 
\ben 
\left[ -\frac{1}{r^2}(\partial_t - i \mu)^2 + \frac{1}{r^3}
  \partial_r \left( r^{5} f_{sl}(r)\partial_r \right)\right]\Phi -
m^2 \Phi - \Phi |\Phi|^2 = 0 \ .
\label{3-6}
\een
By a change of coordinates and a redefinition of fields
\ben
r  \rightarrow  {\rho(r)} = \int_r^\infty \frac{ds}{s^2 f_{sl}^{1/2}(s)}, ~~~~~
\Phi(r,t)  = 
\frac{1}{[r(\rho)]^{\frac{1}{2}}}\left(
  \frac{d\rho}{dr}\right)^{1/2} \Psi (\rho,t)
\label{3-7}
\een
the equation (\ref{3-6}) becomes
\ben
\left[-\partial_t^2 +2i\mu \partial_t \right] \Psi = \cP \Psi - \mu^2
\Psi + 
 G(\rho) |\Psi|^2 \Psi \ .
\label{3-8}
\een
where
\ben
\cP  = -\partial_\rho^2 + V_0(\rho),~~~~~V_0(\rho)  =  m^2 r^2 +
\frac{15 r^8 - 18 r^4 - 1}{4r^2(r^4-1)},~~~~~G(\rho) = \frac{1}{r(\rho)\sqrt{f_{sl}(r(\rho))}}
\label{3-9}
\een
In the new coordinates, near the asymptotic boundary $r = \infty$ we have $\rho \sim 1/r$, while near the tip of the soliton $\rho \sim \rho_\star +\sqrt{r-1}$ where $\rho_\star = 1.311$. As $ \rho \rightarrow 0$ the solution to the linearized version of (\ref{3-8}) behaves as 
\ben
\Psi (\rho,t) \sim \rho^{\alpha} J(t) [1 + O(\rho^2)] + \rho^{1-\alpha} A(t) [1 + O(\rho^2)]~~~~~\alpha = 1/2-\sqrt{m^2 + 4}
\label{3-10}
\een
In standard quantization $J(t)$ is the source, while $A(t)$ is the response, while in alternative quantization they are interchanged. We will discuss the problem in standard quantization - the conclusions in alternative quantization are similar.

We will choose a gauge where the time independent solution is real.  It is shown in \cite{holo-kz2} that with these boundary conditions the operator $\cP$ has a positive spectrum. This means that the operator $\cD = \cP - \mu^2$ develops a {\em zero mode} at some value of the chemical potential $\mu = \mu_{c1}$. For $\mu > \mu_{c1}$ $\cD$ develops a negative eigenvalue. However, now the full non-linear equation has a nontrivial static solution with a nonzero value of the coefficient of $\rho^{\nu + 2}$, but a vanishing coefficient of the $\rho^{1/2-\nu}$ term in (\ref{3-11}).  It is straightforward to show that once there is such a nontrivial solution, it has a lower enegry than the trivial solution. The AdS/CFT correspondence then implies that for $\mu > \mu_{c1}$ the operator $\cO$ which is dual to the bulk scalar field acquires an expectation value. Since this is an operator which is charged under a global $U(1)$ this signifies spontaneous symmetry breaking.

\begin{figure}[!h]
\centering\includegraphics[width=3.0in]{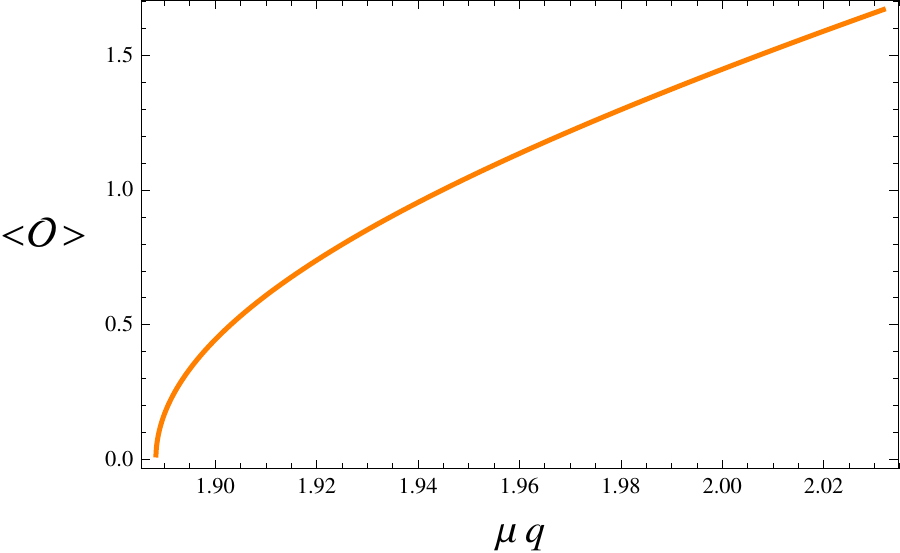}
\caption{Scalar Condensate as a function of $\mu q$. The value of $\mu_{c1}q$ is 1.89. This figure is taken from \cite{holo-kz2}.}
\label{condensate}
\end{figure}

The point $\mu = \mu_{c1}$ is a critical point. A simple analysis near the critical point shows that
\ben
<\cO>|_{J=0} \sim (\mu - \mu_{c1})^{1/2}
\label{3-11}
\een
Figure (\ref{condensate}) shows the result of a numerical (time independent) solution of the nonlinear equation of motion (\ref{3-8}) which is consistent with this behavior.

It is straightforward to derive the critical exponents when the scalar potential in the bulk action is of the form $|\Phi|^{n+1}$. In this case, one has
\ben
<\cO>|_{J=0} \sim (\mu - \mu_{c1})^{1/(n-1)}
\een
Finally the behavior of the order parameter for $J \neq 0$ are
\bea
<\cO>|_{\mu= \mu_{c1}} &  \sim & |J|~~~~~\mu \neq \mu_{c1} \nonumber \\
<\cO>|_{\mu= \mu_{c1}} &  \sim & |J|^{1/n}~~~~~\mu = \mu_{c1} 
\label{3-12}
\eea
These are of course mean field exponents.

In the original work of \cite{holosuper} there was no self-interaction of the scalar. In that case the coupling $\lambda$ in (\ref{3-1}) can be of course scaled out. \cite{holosuper} studied the problem in a probe approximation where the gravity backreaction is ignored, by solving the coupled scalar-Maxwell equations. The exponents obtained there are the same as our theory with a $|\Phi|^4$ potential. The equilibrium problem with back-reaction has been studied in \cite{holosuper2}.

\subsection{Quantum Quench Dynamics}

We now study the response of this system to a quantum quench performed by tuning the chemical potential $\mu$ to be exactly $\mu_{c1}$ and  imposing time dependent boundary condition with a non-trivial $J(t)$ which interpolates between constant values, crossing the critical point at $J=0$ at some intermediate time.
In the boundary quantum field theory this is a time dependent but spatially homogenous external source for the order parameter field $\cO$.
Consider, e.g. the specific profile
\ben
J(t) = J_0 \tanh (t/\delta t)
\label{3-13}
\een
Clearly, for $t \rightarrow -\infty$ the time evolution of the system is adiabatic. The adiabatic expansion of a solution of the equation of motion (\ref{3-8}) is of the form
 \ben
\Psi(\rho,t) = \Psi^{(0)}(\rho,J(t))+ \epsilon \Psi^{(1)}(\rho,t) +
\epsilon^2 \Psi^{(2)} + \cdots \ , 
\label{3-14}
\een
where $\epsilon$ is the adiabaticity parameter. The function  
$\Psi^{(0)}(\rho,J)$  is the
solution of
\ben
\cD \Psi^{(0)} + G(\rho) |\Psi^{(0)}|^2 \Psi^{(0)} = 0
\label{3-14a}
\een
at $\mu = \mu_{c1}$ 
which is regular at the tip and which has a nonzero source $J$. The leading term in (\ref{3-14}) is obtained by replacing $J \rightarrow J(t)$. From (\ref{3-10}) and (\ref{3-12}) the asymptotic behavior of the solution will be of the form
\ben
\Psi_0(\rho,J(t)) \sim \rho^{\alpha} J(t) [1 + O(\rho^2)] + \rho^{1-\alpha} |J(t)|^{1/3} [1 + O(\rho^2)],
\label{3-15}
\een
The adiabatic expansion now proceeds by replacing $\partial_t
\rightarrow \epsilon \partial_t$ and equating terms order by order in $\epsilon$. 
To lowest order we have the following equations for the real and imaginary parts
of $\Psi^{(1)}$
\ben
\left[ \cD + 3 G(\rho) (\Psi^{(0)})^2 \right] ({\rm Re}\, \Psi^{(1)}) 
=  0  ~~~~~~
\left[ \cD +  G(\rho) (\Psi^{(0)})^2 \right] ({\rm Im}\, \Psi^{(1)}) 
=  2\mu\, \partial_t \Psi^{(0)} 
\label{3-16}
\een
Note that in these equations the time dependence of $J(t)$ should be ignored. To solve these equations, we need to find the Green's function for the operator $ \cD +  G(\rho) (\Psi^{(0)})^2 $. However the operator $\cD$ has a zero mode at $\mu = \mu_{c1}$, so that when $J=0$ this Green's function does not exist. For a small nonzero $J(t)$, it is possible to use (\ref{3-15}) to estimate the $J$ dependence of the Green's function and hence the
 leading adiabatic correction $\Psi^{(1)}$. The result is
\ben
{\rm Im}\,\Psi^{(1)} \sim \frac{{\dot{J}(t)}}{J^{2/3}}  \frac{\partial
  \Psi^{(0)}}{\partial J(t)}  \sim \frac{{\dot{J}(t)}}{J^{4/3}} \ .
\label{3-18}
\een
Thus adiabaticity breaks down when we reach a time $t = t_{adia}$ where
\ben
| \Psi^{(1)} (t_{adia}) | \sim | \Psi^{(0)} (t_{adia}) | \implies {\dot{J}(t_{adia})} \sim J^{5/3} (t_{adia}) \ .
\label{3-19}
\een
As in usual mean field theory the gap in the presence of an external source for $\mu = \mu_{c1}$ is $|J|^{2/3}$. When  the time scale of the quench is much larger than the initial  inverse gap, i.e. $\delta t \gg J_0^{-2/3}$, the breakdown of adiabaticity happens in the regime where we can replace $\tanh (t/\delta t) \rightarrow t/\delta t$ and one finally obtains
\ben
t_{adia} \sim \left( \delta t/{J_0} \right)^{2/5}~~~~~~~~~
<\cO(t_{adia})> \sim [J(t_{adia})]^{1/3} \approx [J_0 (t_{adia}/\delta t)]^{1/3} \sim \left( {J_0}/{\delta t} \right)^{1/5}
\label{3-20}
\een
The time $t_{adia}$ is indeed the Kibble-Zurek time in the problem. In this case $r=1$ in (\ref{1-1}) while $z\nu = 2/3$, which yields $t_{KZ} \sim (\delta t)^{2/5}$.

An adiabatic expansion is a power series expansion in $(\dt)^{-1}$. For $t > t_{adia}$ this power series expansion breaks down. We will now argue, however, that in the critical region a new expansion holds - an expansion in {\em fractional} powers of $(\dt)^{-1}$. To see this we first separate out the source term from the field $\Psi$ 
\ben
\Psi (\rho,t) = \rho^{\alpha}J(t) + \Psis(\rho,t) \ , 
\label {3-21}
\een
The boundary behavior of $\Psis$ then starts with $\rho^{1-\alpha}$.

The determination of the adiabaticity breakdown time suggests the rescalings
\ben
t = (J_0/\dt)^{-2/5} \eta \ ,~~~~~~~~~~~~~~\Psi_s = (J_0/\dt)^{1/5} \chi 
\label{3-22}
\een
The key point is that in the critical region we can approximate
\ben
J(t) \approx J_0 \frac{t}{\dt}= (J_0/\dt)^{3/5} \eta
\label{3-23}
\een
Then the equation (\ref{3-8}) admits an expansion in powers of $(J_0/\dt)^{2/5}$,
\ben
\cD \chi =  (J_0/\dt)^{2/5} \left[2i\mu \partial_\eta \chi - G(\rho)
  |\chi|^2 \chi - \eta (\cD \rho^\alpha) \right] +O((J_0/\dt)^{4/5}) \ .
\label{3-24}
\een 
Let us now perform a spectral decomposition of the field in terms of the eigenfunctions of the operator $\cD$,
\ben
\cD \varphi_n (\rho) = \lambda_n \varphi_n (\rho) \ ,~~~~~n=0,1,\cdots \ ,
\label{3-25}
\een
The eigenvalues $\lambda_n$ are discrete because of regularity conditions at the tip and the boundary conditions at the AdS boundary. Recall that for $\mu = \mu_{c1}$ there is a zero mode which we label by $n=0$, i.e.  $\lambda_0 = 0$. All the higher eigenvalues are positive. The spectral decomoposition is 
\ben
\chi(\rho,\eta) = \sum_n \chi_n (\eta) \varphi_n(\rho) \ ,
\label{3-26}
\een
Rewriting the equation (\ref{3-8}) in terms of the modes $\chi_n$ we get an infinite number of coupled ODE's
\ben
\lambda_n \chi_n = (J_0/\dt)^{2/5} \left[ 2i\mu (\partial_\eta \chi_n) -
  \sum_{n_1n_2n_3} \cC^{n}_{n_1n_2n_3} \chi_{n_3}^\star \chi_{n_2}
  \chi_{n_1} + \cJ_n \eta \right] + O((J_0/\dt)^{4/5}) \ ,
\label{3-27}
\een
where we have defined
\ben
\cJ_n  =  \int d\rho \varphi_n^\star (\rho) (\cD \rho^\alpha)~~~~~
\cC^{n}_{n_1n_2n_3}  =  \int d\rho \varphi_n^\star(\rho)
\varphi_{n_3}^\star(\rho) 
\varphi_{n_2}(\rho) \varphi_{n_1}(\rho) G(\rho) \ .
\label{3-28}
\een
It is clear from (\ref{3-27}) that the zero mode dominates the dynamics for small $(J_0/\dt)$. 
In fact the solution may be written in the form
\ben
\chi_n(\eta) = \delta_{n0} \xi_0(\eta) + (J_0/\dt)^{2/5} \xi_n + O((J_0/\dt)^{4/5}) \ .
\label{3-29}
\een
The zero mode $\xi_0$ satisfies a $z=2$ Landau-Ginsburg equation 
\ben
-2i\mu \partial_\eta \xi_0 + \cC^0_{000} |\xi_0|^2 \xi_0 +\cJ_0 \eta =
0 \ .
\label{3-30}
\een
Reverting back to the original variables we therefore have
\ben
\Psi_s (\rho,t,\dt) = (J_0/\dt)^{1/5} \Psi_s (\rho,t (J_0/\dt)^{2/5},1) \ ,
\label{3-31}
\een
which implies a scaling solution for the order parameter with $z=2$
\ben
\langle \cO(t,\dt)\rangle  = (J_0/\dt)^{1/5} \langle \cO ((J_0/\dt)^{2/5}t,1)\rangle   \ .
\label{3-32}
\een
Note that the effective Landau-Ginzburg equation (\ref{3-30}) is not dissipative because the first order time derivative is multiplied by a purely imaginary constant. In fact, in the absence of a source term the quantity $\frac{1}{2} (|\xi_0|^2)^2$ is independent of time. 

The effect of backreaction of the gauge field and bulk gravity in the critical region has been studied in \cite{holo-kz4}. The small oscillation operators for the gauge field and the metric components do not have zero modes. However the scalar zero mode drives non-adiabatic evolution of these as well. Once again there is an expansion in fractional powers of $\dt$ leading to scaling functions for the expectation value of the current and the energy momentum tensors
\bea
<J_\mu (t,\dt) > & \sim & (J_0/\dt)^{2/5} <J_\mu (t (J_0/\dt)^{2/5},1 ) > \nonumber \\
<T_{\mu\nu} (t,\dt) > & \sim & (J_0/\dt)^{2/5} <T_{\mu\nu} (t (J_0/\dt)^{2/5},1)>
\label{3-33}
\eea

\begin{figure}[!h]
\centering\includegraphics[width=3.0in]{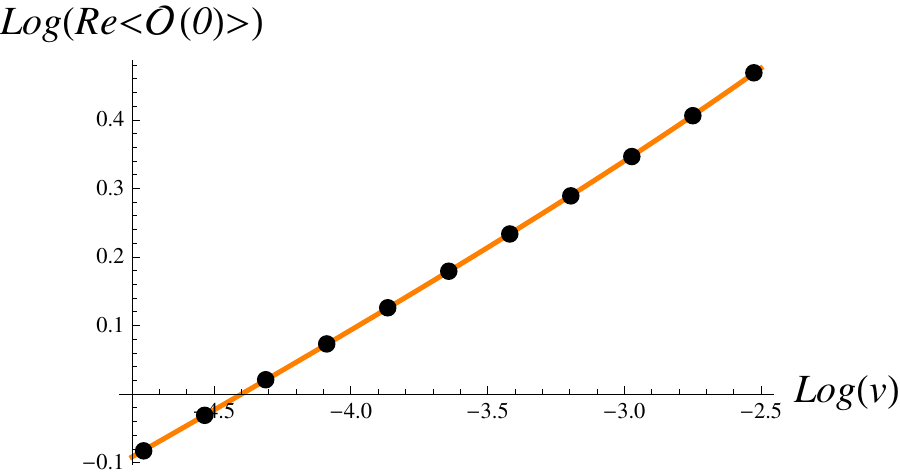}
\caption{(Colour online) Real part of the dual operator at $t=0$ as a function of the logarithm
 of the quench rate $v = dt^{-1}$. This figure is taken from \cite{holo-kz2}.} 
\label{soliton-quench}
\end{figure}

Figure (\ref{soliton-quench}) shows the result of a direct numerical solution of the full nonlinear scalar equation. Here the logarithm of the real part of the expectation value of the dual operator (extracted from the fall off of the solution) at $t=0$ is plotted as a function of $-\log (\dt)$ for a quench with $J(t) = J_0 \tanh (t/\dt)$. The straight line is the best fit : $0.794 + 0.490/(\log(\dt))-0.206 \log(\dt)$, consistent with our analytic result $<\cO> \sim \dt^{-1/5}$.

The features of this model which lead to a scaling solution are common to all the models studied in \cite{holo-kz1}-\cite{holo-kz4}. In each case, a zero mode appears at the critical point. During the times when the system passes in the vicinity of the critical point, a new large-$\dt$ expansion appears in fractional powers of $\dt$. In the lowest order of this expansion the zero mode completely dominates the dynamics, and this directly leads to a scaling solution.

The naive argument which leads to Kibble Zurek scaling presented in the introduction has two features which requires justification. The first is the assumption that the system becomes diabatic after $t=t_{KZ}$. Our holographic considerations of course do not involve any such assumption. The second assumption is that the instantaneous correlation length at the Kibble Zurek time $t_{KZ}$ is the only scale in the problem - i.e. a  decoupling of scales in the dynamics. In a static situation, decoupling of scales and the dominance of a few scales in the critical region is understood in terms of RG. This is not something which is well understood in the dynamical situation. Our holographic models may have something to teach in this regard. In AdS/CFT the radial coordinate plays the role of a scale in the boundary theory. The mode decomposition of the bulk field in terms of eigenfunctions of $\cD$ is in some sense a decomposition in terms of scale. The dominance of the zero mode in the dynamics therefore indicates a decoupling of scales. Hopefully this insight will be useful in understanding the issue purely in field theory without recourse to holography.

\section{Fast Quenches}

The previous section dealt with quantum quench in holographic models at a rate which is much {\em slower} than the mass scales in the theory. We now turn to scaling relations found in \cite{numer} and \cite{fastQ} in holographic quenches (both thermal and quantum) at rates which are {\em fast} compared to all physical mass scales. The quench was implemented in a way similar to \cite{bh3} and \cite{holo-kz1} - \cite{holo-kz4}, by imposing a time dependent but spatailly homogeneous boundary condition on a bulk scalar field with different masses with $m^2 < 0$, so that this corresponds to deformations of a CFT with relevant  operators with different conformal dimensions $\Delta$. In \cite{numer} the response soon after the quench was calculated numerically using standard techniques of holographic renormalization. For small $\dt$ the results were consistent with the universal scaling formulae (\ref{1-11}). In \cite{fastQ} an analytic understanding of this scaling was achieved. In the regime of small $\dt$ bulk causality implies that the response is determined by a small region near the boundary. It turns out that in this region the non-linearities in the bulk equations are not important, and the scaling property follows simply.

The result (\ref{1-11}) appears puzzling at first sight, since it implies that for $\Delta > d/2$ the response diverges as $\dt \rightarrow 0$, whereas instanatneous quenches, as discussed in \cite{cc2,cc3} appear to make sense, at least in low dimensions. However, the holographic results are for {\em renormalized} quantities and are therefore makes sense for quench rates which are {\em slow compared to the UV cutoff scale}, whereas an instantaneous quench strictly means a quench rate which is {\em fast compared to all scales}. In this section we will show that this universal scaling is in fact a general property of {\em any quantum field theory}, regardless of holography and holds for intermediate quench rates which lie inbetween abrupt quenches and the slow quenches discussed in the previous section. In the following we will summarize some salient aspects of \cite{dgm1,dgm2,dgm3,dgm4}. 

\subsection{The General Result}

Consider a relativistic quantum field theory in $d$ space-time dimension described by an action 
\ben
S = S_{CFT} - \int dt \int d^{d-1}x ~\lambda (t) \cO (\vx,t)
\label{4-1}
\een
where $\cO$ is a relevant operator with conformal dimension $\Delta$. Here $S_{CFT}$ is the action of a CFT in the UV. The time dependence of the coupling $\lambda (t)$ will be taken to be of the form
\ben
\lambda (t) = \lambda_0 + (\delta \lambda) F(t/\dt)
\label{4-2}
\een
Where the function $F(x)$ vanishes for $x < 0$ and is equal to one for $x > 1$. The function is non-trivial for $0 \leq x \leq 1$ with excursions of of order one. The function is also continuous and integrable.

\begin{figure}[!h]
\centering\includegraphics[width=3.5in]{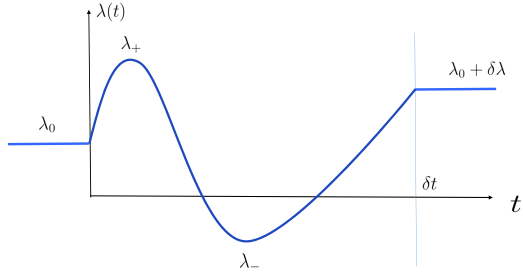}
\caption{Profile of coupling $\lambda(t)$}
\label{profile3}
\end{figure}

The Heisenberg picture state is the ground state of the hamiltonian at early times, i.e. of the hamiltonian of a conformal field theory deformed by  a relevant operator with constant coupling $\lambda_0$. This theory thus has a gap $m_g$ which is of order $\lambda_0^{\frac{1}{d-\Delta}}$.

Let us start computing the quantity $<\cO(\vx,t)>$ in perturbation theory. The first few terms are given by
\bea
&&\langle \calo_\Delta(\vx,t)\rangle^{\rm ren} =
\langle \calo_\Delta(\vx,t) \rangle^{\rm ren}_{\lambda_0}- \delta \lambda \int_0^t dt^\prime \int d^{d-1}x^\prime~ F(t^\prime/\dt)\,  G_{R,\lambda_0} (\vx,t; \vx^\prime, t^\prime) 
\label{4-3}\\
&&\qquad\qquad\qquad+ \frac{\delta\lambda^2}{2} \int d^{d-1} x^\prime d t^\prime ~F(t^\prime/\dt)\,\int d^{d-1} x^{\prime\prime} d t^{\prime\prime}\, F(t^{\prime\prime}/\dt) \ K_{\lambda_0}(\vx, \vx', \vx'', t,t',t'')
+\cdots,
\nonumber
 \eea
where
\begin{eqnarray}
G_{R,\lambda_0} (\vx,t;\vx^\prime,t^\prime) = i \theta(t)\ \langle\,  [\calo_\Delta (\vx,t) , \calo_\Delta (\vx^\prime,t^\prime) ]\, \rangle_{\lambda_0}\,.
\end{eqnarray}
is the retarded correlator while $K(x,x',x'',t,t',t'')$ is a three point function. These are objects in the initial theory which has space-time translation invariance, so that $G_{R,\lambda_0} (\vx,t;\vx^\prime,t^\prime) = G_{R,\lambda_0} (\vx -\vx^\prime, t -t^\prime)$. Since the couplings depend only on time, the one point function \\
$<\calo (\vx,t)>$ is independent of $\vx$.  We will assume that the field theory can be renormalized in a standard fashion. All quantities in (\ref{4-3}) are {\em renormalized quantities} - the UV cutoff has been taken to be infinity. In the next subsection we will consider some specific examples and comment on the nature of renormalization involved.

\begin{figure}[!h]
\centering\includegraphics[width=3.5in]{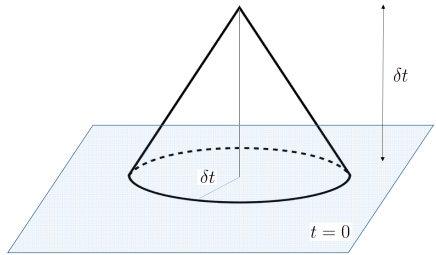}
\caption{Region of integration in (\ref{4-4})}
\label{causal}
\end{figure}

Consider the first correction in (\ref{4-3}) 
\ben
\delta \lambda \int_0^t dt^\prime \int d^{d-1}x^\prime~ F(t^\prime/\dt)\,  G_{R,\lambda_0} (\vx - \vx^\prime, t - t^\prime) 
\label{4-4}
\een
Note that $G_{R,\lambda_0}$ is a causal propagator which vanishes outside the past light cone of the point $(\vx,t)$. This means
while the integration over $\vx^\prime$ has been written over the entire space, only the region $|\vx -\vx'| \leq t$ has a non-zero contribution. Now suppose we want to calculate the response at $t=\dt$. Then both the time and space intervals involved in the integral are at most of size $\dt$, as shown in Figure (\ref{causal}).

Consider now the {\em fast quench} regime. This means
\ben
\dt \ll (\delta \lambda)^{-\frac{1}{d-\Delta}}, (\lambda_{0,\pm})^{-\frac{1}{d-\Delta}}
\label{4-5}
\een
Then the region of integration is small compared to all other physical length scales in the problem, in particular the correlation length of the theory with coupling $\lambda_0$. In this regime the correlators of the theory are indistinguishable from the correlators of the UV conformal field theory,
\ben
G_{R,\lambda_0} (\vx - \vx^\prime, t - t^\prime) \approx G_{R,CFT} (\vx - \vx^\prime, t - t^\prime)~~~~~{\rm for}~~~~~
|\vx - \vx^\prime |, |t - t^\prime | \ll (\lambda_{0})^{-\frac{1}{d-\Delta}}
\label{4-6}
\een
This means that the only scale which appears in the integral is $\dt$. To leading order, the same would be true for the successive terms in the expansion. Thus the response can be written as
\ben
\langle \calo_\Delta(t)\rangle^{\rm ren} -
\langle \calo_\Delta(t) \rangle^{\rm ren}_{\lambda_0}
=(\delta \lambda)^{-\Delta} \left[ b_1 (t/\dt) g + b_2 (t/\dt) g^2 + \cdots \right]
\label{4-7}
\een
where we have defined a dimensionless coupling
\ben
g \equiv (\delta \lambda) (\dt)^{d-\Delta}
\label{4-8}
\een
and $b_i(t/\dt)$ some functions of $t/\dt$.
In other words, in the fast quench limit the dimensionless coupling is small and the answer is given reliably by perturbation theory. The leading contribution is clearly
\ben
\langle \calo_\Delta(t)\rangle^{\rm ren} -
\langle \calo_\Delta(t) \rangle^{\rm ren}_{\lambda_0} \sim (\dt)^{d-2\Delta}~~~~{\rm for}~~~~~~t \sim \dt
\label{4-9}
\een
which is exactly what was found in holographic models.

A similar argument reproduces the scaling for the energy density in (\ref{1-11}), as would also be required by the Ward identity
\ben
\frac{d \cE}{dt} = - \frac{d \lambda(t)}{dt} <\cO>
\een
and similar scaling should hold for other quantities as well. Note that the scaling property follows entirely from the properties of the UV conformal field theory - the IR behavior is not important. Basically this is because we are looking at the short time response, i.e. around the time when the quench is getting over.

\section{Free field quenches}

A lot of insight into this problem can be in fact obtained by looking at  field theories with time dependent parameters whose time evolution can be exactly solved. In \cite{dgm1}-\cite{dgm4} we considered free bosonic and fermionic field theories with time dependent masses,
\bea
S & = & -\int dt\int d^{d-1}x~\frac{1}{2}[ (\partial \phi)^2 + m^2(t) \phi^2 ] 
\label{4-11} \\
S & = &\int dt\int d^{d-1}x~{\bar \psi}[i\gamma^\mu \partial_\mu + M(t)] \psi
\label{4-12}
\eea
The bosonic theory can be solved for mass profiles of the form
\bea
m^2(t) & = & \frac{1}{2} m^2[A + B \tanh(t/\dt) ]\label{4-13} \\
m^2(t) & = & m_0^2 + \frac{m_1^2}{\cosh^2 (t/dt)}
\label{4-14}
\eea
while the fermionic theory will be solved for profiles
\ben
M(t) = M(C + D \tanh (t/\dt))
\label{4-15}
\een
In (\ref{4-13}) - (\ref{4-15}) $A,B,C,D,m_0$ and $m_1$ are arbitrary constants. Clearly at early and late times the masses asymptote to constant values. By choosing these constants appropriately the time dependent masses can be made to vanish at some intermediate time. A vanishing mass is a quantum critical point.

For example, when $A = - B = 1/2$ the scalar field mass in (\ref{4-11}) starts with $m$ and approaches zero as $t \rightarrow \infty$. This is like the setup of \cite{cc2,cc3} where the quench is from a massive theory to a massless theory. A choice $m_0=0$ describes a quench from a massless theory to a massless theory.
A choice of $C=0$ and $D=1$ corresponds to a fermion mass which goes from $M$ to $-M$, passing through zero at $t=0$. However since the sign of a fermion mass is not physical, this is a quench from a massive phase to a massive phase, passing through a critical theory, pretty much like the setup used in discussing Kibble Zurek scaling. A protocol which can be used to discuss Kibble Zurek physics for scalars is given by $m_1^2 = -m_0^2$ in (\ref{4-14}) - once again the quench is from a massive phase to another massive phase passing through a critical point at $t=0$.

These models can be solved for arbitrary $\dt$. This means that the time evolution of the field operator $\phi (\vx,t)$ or $\psi (\vx,t)$ can be exactly determined.

\subsection{Scalar Quenches} We now outline the main steps involved in calculating the response to quenches in scalar field theories. The steps for fermions are very similar, albeit a bit more involved. All details can be found in \cite{dgm2}.

Consider for example the bosonic theory with a mass profile (\ref{4-13}). The mode decomposition of the field is
\cite{birell-davies}
\ben
\phi= \int\!\! \frac{d^{d-1}k}{(2\pi)^{(d-1)/2}} \ \left( a_\vk\, u_\vk + a^\dagger_\vk\, u^*_\vk\right)\,,\qquad
{\rm where}\ \ \ [a_\vk , a^\dagger_{\vk^\prime} ] = \delta^{d-1}(\vk - \vk^\prime)\,.
\label{5-1}
\een
The functions $u_\vk (\vx,t)$ are solutions of the equations of motion. Since we are interested in studying quenches starting from a vacuum state, we will choose  them to obey the initial condition that at early times they are plane waves with the appropriate mass. The solution is then
\begin{eqnarray}
u_\vk & = & \frac{1}{\sqrt{2 \omega_{in}}} \exp(i\vk\cdot\vec{x}-i\omega_+ t - i\omega_- \dt \log (2 \cosh t/\dt)) \times \nonumber \\
& &\qquad_2F_1 \left( 1+ i \omega_- \dt, i \omega_- \dt; 1 - i \omega_{in} \dt; \frac{1+\tanh(t/\dt)}{2} \right)\,,
\label{5-2}
\end{eqnarray}
where
\ben
\omega_{in}  =  \sqrt{\vk^2+m^2 (A-B)}~~~~~~
\omega_{out}   =  \sqrt{\vk^2+m^2 (A+B)}~~~~~~
\omega_{\pm}   =  (\omega_{out}\pm\omega_{in})/2
\label{5-3}
\een
It may be easily checked that
\ben
{\rm Lim}_{t \rightarrow -\infty} u_\vk (\vx,t) = \frac{1}{\sqrt{2\omega_{in}}}~{\rm exp} \left[ i (\vk \cdot \vx) - i\omega_{in} t \right]
\label{5-4}
\een
There are similar mode decompositions for the mass profile (\ref{4-14}) and for the fermions. These can be found in \cite{dgm2}.

The Heisenberg picture state in the problem is given by the "in" vacuum, 
\ben
a_\vk |in,0\rangle=0
\label{5-5}
\een
The response of the system is then given by
\ben
\langle\phi^2\rangle\equiv
\langle in,0|\phi^2|in,0\rangle = \frac{1}{2(2\pi)^{d-1}}\int \frac{d^{d-1}k}{\omega_{in}}\, |_2F_1|^2\,.
\label{5-6}
\een
This is of course divergent : we need a UV cutoff and a renormalization procedure to make sense of the result. Even though we are dealing with a free field, renormalization is non-trivial since the mass is time dependent. The problem is similar to that of a free quantum field in curved space-time. There the counter-terms needed to render the theory finite depend on curvature invariants. In this case, a systematic way of finding the necessary counterterms is to put the theory on a curved background metric and think of the mass as another scalar field. The idea is then to write down all possible invariant counterterms which involve the metric, the mass field and their derivatives with powers of the UV cutoff $\Lambda$ , retaining only terms which diverge when $\Lambda \rightarrow \infty$, thus yielding a counterterm action $S_{ct}$. This would lead to a renormalized action $S_0 + S_{ct}$ where all UV divergences have been cancelled. The renormalized response and the energy momentum tensors can be then obtained by taking appropriate derivatives of the path integral which follows from this renormalized action.

However there is a simpler and perhaps more intuitive way of obtaining the counterterms. This involves subtracting the adiabatic expansion from (\ref{5-6}).  Let us express the modes in a form
\ben
u_\vk = \frac{1}{\sqrt{2\, \Omega_k(t)}} \exp \left(i \, \vk\cdot\vec{x} -i \int^t \Omega_k(t') dt' \right)\,.
\label{5-7}
\een
This solves the equation of motion provided
\ben
\Omega_k^2 = \omega_k^2 - \frac{1}{2}  \frac{\partial_t^2{\Omega}_k}{\ \Omega_k} + \frac{3}{4} \left(\frac{\partial_t{\Omega}_k}{\Omega_k} \right)^2 \,,
\qquad{\rm with}\ \ \omega_k^2= k^2 + m^2(t)\,. \label{5-8}
\een
The adiabatic expansion is obtained by solving (\ref{5-8}) in an expansion in time derivatives. To the lowest non-trivial order the answer is
\ben
\Omega_k (t) = \omega_k (t)  -\frac{1}{2} \left(\frac{\ddot{\omega}_k}{\omega_k} -\frac{3 \dot{\omega}_k^2}{2\omega_k^2} \right) + \cdots
\label{5-9}
\een
Finally the response in this expansion is
\ben
\langle\phi^2\rangle^{adiabatic}
 = \frac{1}{2(2\pi)^{d-1}}\int \frac{d^{d-1}k}{\omega_{in}}\, \frac{1}{\Omega_k (t)}.
\label{5-11}
\een
It turns out that the UV divergent terms in (\ref{5-11}) exactly agree with those which appear in (\ref{5-6}). Thus the quantity 
\ben
\langle\phi^2\rangle^{{\rm ren}} = \langle\phi^2\rangle - \langle\phi^2\rangle^{adiabatic}
\label{5-12}
\een
is finite.

It may appear strange that the adiabatic expansion, which is valid for {\em slow} changes of parameters, has the necessary counterterms for {\em arbitrary} rates of changes of $m^2(t)$. The reason is simple. The way we extracted the divergent pieces from the adiabatic expansion assumed that the UV cutoff $\Lambda$ is much higher than all other scales in the problem, including the scale of change of the mass, the quench rate. So for any quench rate which is {\em slow} compared $\Lambda$, the contributions which come from momenta near the cutoff do not care if the quench rate is slow or fast compared to some other IR mass scale.

For general interacting theories we still expect that the counterterms can be read off from an adiabatic expansion, though this would be much more complicated.

\subsection{The fast quench regime} 

The renormalized expectation value (\ref{5-12}) can be evaluated by performing the momentum integral numerically : the results are discussed in detail in \cite{dgm1} and \cite{dgm2}. However when the quench rate is {\em fast}, i.e. $m \dt \ll 1$ analytic expressions can be derived in powers of $(m\dt)$. Here we quote the results for $3 \leq d \leq 9$ in Table (\ref{fastresults})

\begin{table}[!h]
\caption{Fast Quench Expressions for $<\phi^2>_{\rm ren}$.}\label{fastresults}
\centering
\begin{tabular}{|c||c|}
\hline
{\rm odd}~~$d \geq 5$ & $\langle\phi^2\rangle_{ren} = (-1)^{\frac{d-1}{2}} \frac{\pi}{2^{d-2}}\, \partial^{d-4}_t m^2(t) + O(\dt^{6-d})$ \\ \hline
{\rm even}~~$d \geq 6$ & $(-1)^{d/2} \log(\mu \dt)\, \frac{\partial^{d-4}_t m^2(t)}{2^{d-3}} + \cdots $\\ \hline
$d=4$ & $\frac{m^2}{4} ( 1 + \tanh(t/dt)) \log (\mu \dt) + \phi_2 (t) + O(\dt^2)$ \\ \hline
$d=3$ &$ -\frac{m}{4\pi}  - \frac{m^2\dt}{16}\,  \log \left( \frac{1-\tanh t/\dt}{2} \right) + O(\dt^3)$ \\ \hline
\end{tabular}
\end{table}
Note that in even dimensions we have logarithmic enhancements and we need to introduce a renormalization scale $\mu$. The function $\phi_2 (t)$ is a known function of time, which is given in equation (2.45) of \cite{dgm2}. Since the conformal dimension of the quenched operator is $\Delta = d-2$ it is clear that for $ d \geq 4$ these results are consistent with (\ref{1-11}) upto logarithmic enhancements in even dimensions, which was also observed in the holographic calculations of \cite{numer} and \cite{fastQ}.  For $d=3$ the result is finite in the limit $\dt \rightarrow 0$ : the leading correction then obeys the scaling law.

\begin{figure}[!h]
\centering\includegraphics[width=4.5in]{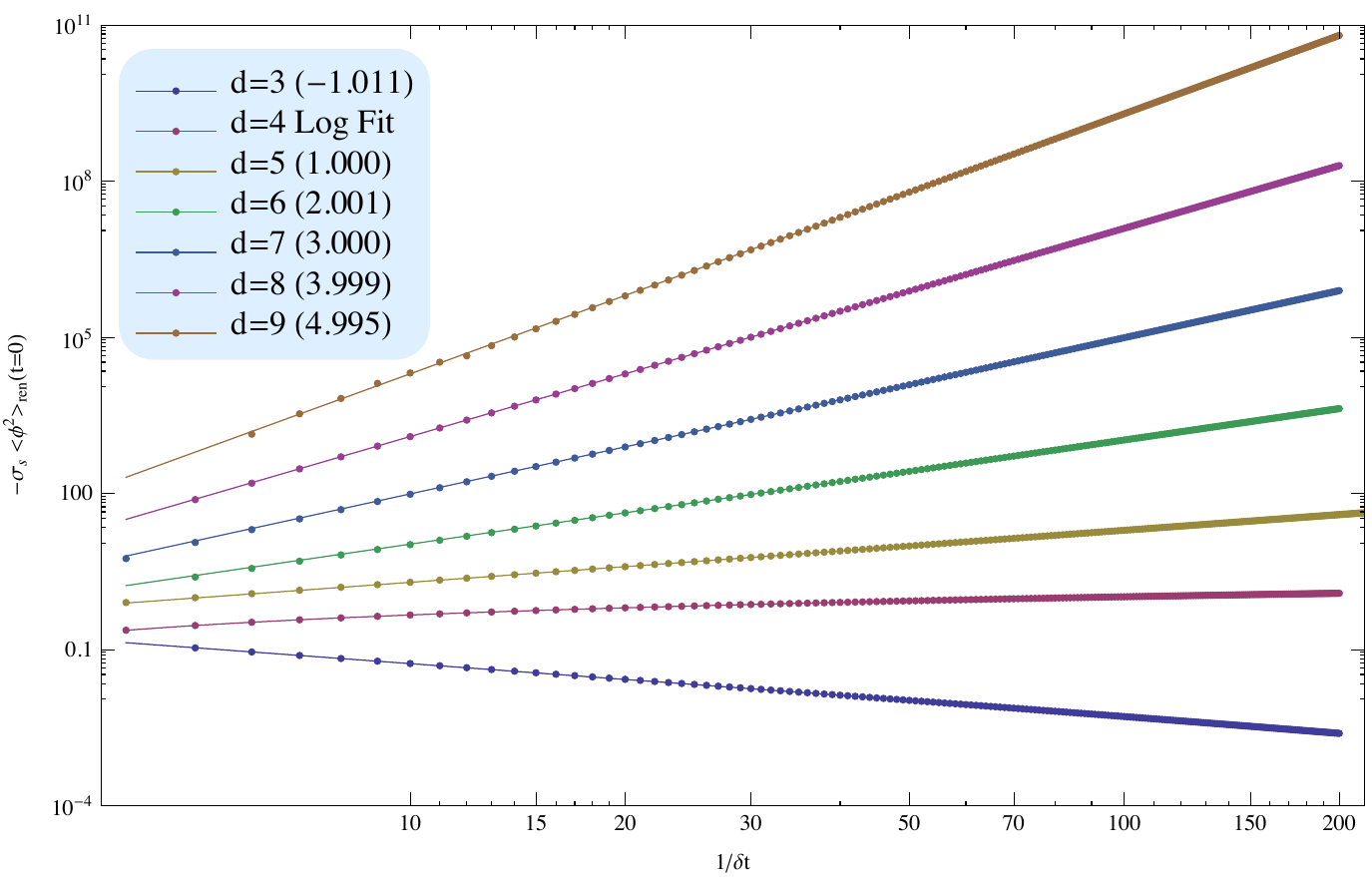}
\caption{(Colour online) Expectation value $\langle\phi^2\rangle_{ren}(t=0)$ as a function of the quench times $\dt$ for spacetime dimensions from $d=3$ to 9. Note that in the plot, the expectation values are multiplied by the numerical factor: $\sigma_s=\frac{2(2\pi)^{d-1}}{\Omega_{d-2}}$.
The slope of the linear fit in each case is shown in the brackets beside the labels.
The results  support the power law scaling $\langle\phi^2\rangle_{ren} \sim \dt^{4-d}$. This figure is taken from \cite{dgm2}.} 
\label{fast-scaling-scalar}
\end{figure}

Figure (\ref{fast-scaling-scalar}) shows the result for a numerical evaluation of the exact integrals involved in equations (\ref{5-6}) and (\ref{5-12}) at $t=0$ as a function of the quench rate $1/\dt$, for various dimensions. The renormalized expectation value at $t=0$ is in excellent agreement with the expected scaling behavior.
 
For other solvable quench protocols for scalars and fermions similar analytic expressions can be obtained in the fast quench regime.
It is also possible to verify the scaling relations for the energy momentum tensor, and similar relations for higher spin currents.

\subsection{From Fast quench scaling to Kibble-Zurek}

We have seen that different universal scalings hold in different quench rate regimes : Kibble Zurek for slow quenches and the more recently discovered novel scaling for fast smooth quenches. We now investigate how one passes from the slow regime to the fast regime. In particular we would like to know if there is a phase transition.

It turns out that both these scaling regimes appear in suitable free field quenches. This has been investigated in detail in \cite{dgm4}. For scalar quenches we need to use (\ref{4-14}) with $m_1^2 = -m_0^2 > 0$ and for fermionic quenches we can use (\ref{4-15}). The slow quench regime should be then described by $m_0\dt \gg 1$ or $M\dt \gg 1$, while the fast quench regime would be as discussed in the previous subsection, $m_0 \dt \ll 1$ or $M \dt \ll 1$. For scalar quenches, we have been able to obtain some analytic understanding of the origin of Kibble Zurek scaling in the slow quench regime. 

As an example, consider a quench protocol (\ref{4-15}) in the fermionic theory. In the fast quench regime $M\dt \gg 1$ the analytic result for the renormalized expectation value of  $<{\bar \psi}\psi>$ is given by
\ben
\langle \bar{\psi}\psi \rangle_{ren}  =   (-1)^{\frac{d}{2}-1}\, \frac{\log(\mu\dt)}{2^{d-2}\,\sigma_f}\, \partial_t^{d-2} m(t/\dt) + O(\dt^{2-d})\qquad {\rm for}\ d\ge4\,,
\label{5-13}
\een
while the expectation from Kibble-Zurek scaling in the slow regime $M\dt \gg 1$ is 
\ben
\langle \bar{\psi}\psi \rangle_{ren} \sim \left(\frac{M}{\dt} \right)^{\frac{d-1}{2}} 
\label{5-14}
\een

\begin{figure}[!h]
\centering\includegraphics[width=4.0in]{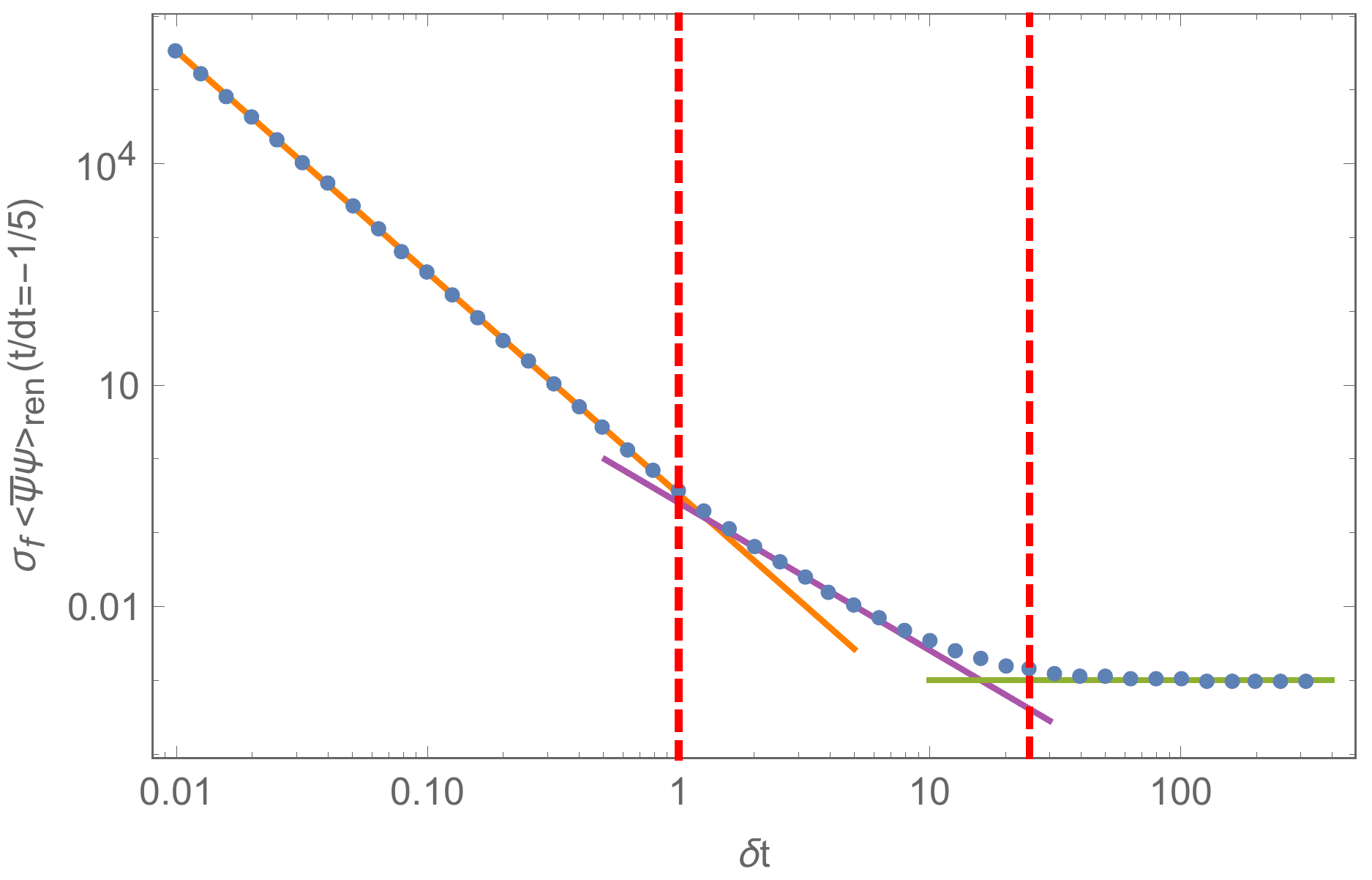}
\caption{(Colour online)
Expectation value at fixed $t/\dt=-1/5$ as a function of $\dt$ for a fermionic quench with $d=5$ and $M=1$. The solid orange line is the analytic leading contribution (\ref{5-13}) for fast quenches; the solid purple line is a linear best fit and agrees with the KZ scaling, $<{\bar{\psi}\psi}>_{ren} \sim \dt^{-2}$; and the solid green line shows the adiabatic value for a fixed mass. As a guide to the eye, the dashed red lines show $\dt=1$ and $\dt=25$, which correspond to the transition regions. This figure is taken from \cite{dgm4}. }
\label{fermionic-fast-slow}
\end{figure}

Figure (\ref{fermionic-fast-slow}) shows a typical result for a quench in the fermionic theory. The quantity which is plotted is the {\em renormalized} expectation value of $<{\bar \psi}\psi>$ measured at a time just prior to the critical point, at $t = -1/5 \dt$. The result clearly shows three regimes. For small $\dt$ the points lie right on top of the analytic answer for fast scaling (\ref{5-13}), shown in solid orange. For $M\dt \sim 1$ there is a cross-over to a Kibble Zurek type of scaling. The purple line is the expected result (\ref{5-14}). Finally for much larger $M\dt$ the dependence on $\dt$ saturates.  This is in fact a range of quench rates for which the time $t = -1/5 \dt$ is still in the adiabatic regime. The points lie on top of the green line which is the expectation value for a constant mass equal to $M(-1/5 \dt)$. 

Our results show that the passage between these three regimes is completely smooth.

\subsection{Instantaneous and Smooth Quenches}

So far we have discussed {\em renormalized} quantities. These are useful to discuss physics at energy scales much lower than the UV cutoff. Thus the quench rates are always much lower than the UV scale. On the other hand, an {\em instantaneous} quench - which is a sudden change of the hamiltonian - involves quench rates which are fast compared to {\em all} scales, even the UV scale. Of course a strictly sudden change is unphysical. For sufficiently fast quenches suitable IR quantities should behave in a way similar to what one would get from instantaneous quenches. The considerations of \cite{cc2,cc3} should be valid for such quantities. Our aim now is to understand the relationship between instantaneous quenches and the fast quenches discussed in the previous subsections.

One way to address these issues is to look at lattice models. Recently we have found suitable solvable quench protocols in interesting lattice models in $1+1$ as well as in $2+1$ dimensions \cite{ddgms}. Here we will continue to discuss continuum theories, but consider {\em UV finite} objects at finite quench rates so that the role of possible ambiguities in the renormalization procedure is absent. The questions we want to understand are (i) Is there a universal scaling for such quantities ? (ii) how do instantaneous quench results compare with those of finite but very fast quench rates ? Our discussion will be in free scalar field theory, but we will draw some lessons for general field theories as well. We will discuss a quench from a massive theory to a massless theory. The following two subsections summarize results of \cite{dgm3}.

\subsubsection{Correlation Functions}

One class of UV finite quantities are equal time correlation functions at finite spatial separations of magnitude $r$
\ben
C(t,r)_{smooth} = <0,in|\phi (\vx,t) \phi(0,t) |0,in>
\label{6-1}
\een
The correlator (\ref{6-1}) may be expressed in terms of Bogoliubov coefficients $\alpha_\vk$ and $\beta_\vk$ which relate the "in" modes we have been using, and "out" modes which are plane waves at late times. The result is
\ben
C(t,r)_{smooth} = \frac1{\sigma_c\,r^{\frac{d-3}{2}}} \int dk \, \,  k^{\frac{d-3}{2}} \,J_{\frac{d-3}{2}}(k r) \, \Big\{ |\alpha_\vk|^2 +|\beta_\vk|^2+ \alpha_\vk \beta^\star_{\vk}~e^{2 i k t}+
\alpha^\star_\vk \beta_\vk ~e^{-2 i k t}  \Big\}
\label{6-2}
\een
where $\sigma_c=2^{\frac{d+1}{2}}\pi^{\frac{d-1}{2}}$ and the Bogoliubov coefficients are given by 
\ben
\alpha_\vk  = \sqrt{\frac{\omega_{out}}{\omega_{in}}} \, \frac{\Gamma (1-i\omega_{in}\delta t)\Gamma(-i\omega_{out}\delta t)}{\Gamma(-i\omega_+\delta t)\Gamma(1-i\omega_+\delta t)} ~~~~~
\beta_\vk  =  \sqrt{\frac{\omega_{out}}{\omega_{in}}} \, \frac{\Gamma (1-i\omega_{in}\delta t)\Gamma(i\omega_{out}\delta t)}{\Gamma(i\omega_-\delta t)\Gamma(1+i\omega_-\delta t)}
\label{6-3}
\een
The correlator for an instantaneous quench, $C(t,r)_{instant}$ is given by the expression (\ref{6-2}) where the Bogoliubov coefficients are now given by
\ben
\alpha_\vk^{instant} = \frac{\omega_+}{\sqrt{\omega_{in}\omega_{out}}} \qquad {\rm and} \qquad    \beta_\vk^{instant} = \frac{\omega_-}{\sqrt{\omega_{in}\omega_{out}}}.
\label{6-4}
\een
These are in fact the limit of (\ref{6-3}) when $\omega_i \dt \ll 1$ for $i=\pm,in,out$. This simply reflects the fact that an instantaneous quench has a rate $\dt^{-1}$ which is large compared to all the momenta in the problem. 

\begin{figure}[!h]
\centering\includegraphics[width=3.0in]{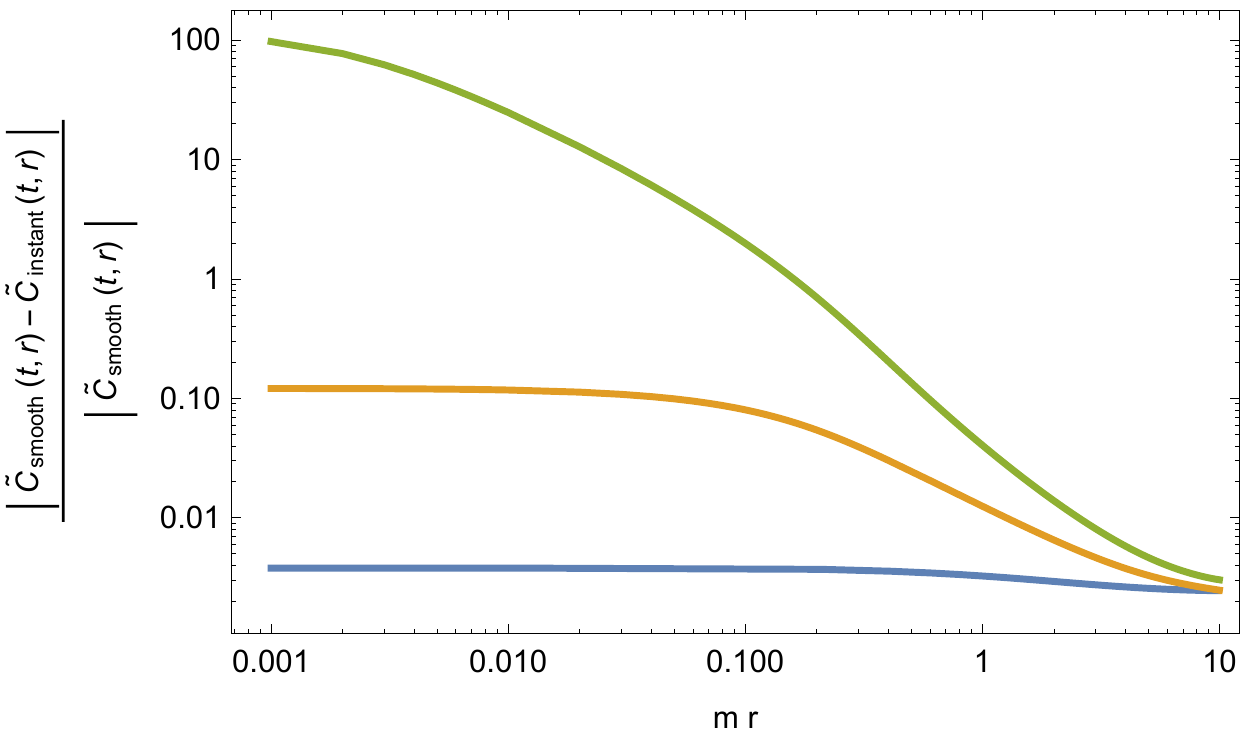}
\caption{(Colour online)Difference between the late-time correlators for smooth and instantaneous quenches as a function of the separation distance $r$. The blue line corresponds to the $d=3$ case while the yellow one belongs to $d=5$ and $d=7$ is shown in green. We are using $m t=10$ with $m \dt=1/20$. In $d=3$ and $d=5$, the difference remains small for any value of $r$, while in $d=7$, it seems to diverge as $r\to0$. This figure is taken from \cite{dgm3}}
\label{late-time}
\end{figure}

One would nevertheless expect that at late times and for large enough separations $C(t,r)_{smooth}$ and $C(t,r)_{instant}$ should not differ much. This is what is found in \cite{dgm3}. To make this comparison it is conveninent to subtract the correlator with a {\em fixed} mass equal to the mass at late times - in this case this is simply the zero mass correlator 
\ben
C(t,r)_{const} =  \frac1{\sigma_c\,r^{\frac{d-3}{2}}} \int
dk\,\,k^{\frac{d-3}{2}}\, J_{\frac{d-3}{2}}(k r)
\label{6-5}
\een
The subtracted correlators are denoted by ${\tilde{C}}(t,r)_{smooth,instant}$
Figure (\ref{late-time}) shows the results for the difference of the smooth and instantaneous correlator normalized by ${\tilde{C}}(t,r)_{smooth}$ as a function of the spatial separations at late times. As one expects, for large $mr$ the difference goes to zero. This reflects the fact that for large $mr$ the momenta which contribute to the integrals are small compared to $\dt^{-1}$ so that the Bogoliubov coefficients agree.  One might have thought that this difference should be small even for finite $mr$ at very late times. However the results show that this is clearly not the case, except for low space-time dimensions.

\begin{figure}[!h]
\centering\includegraphics[width=4.5in]{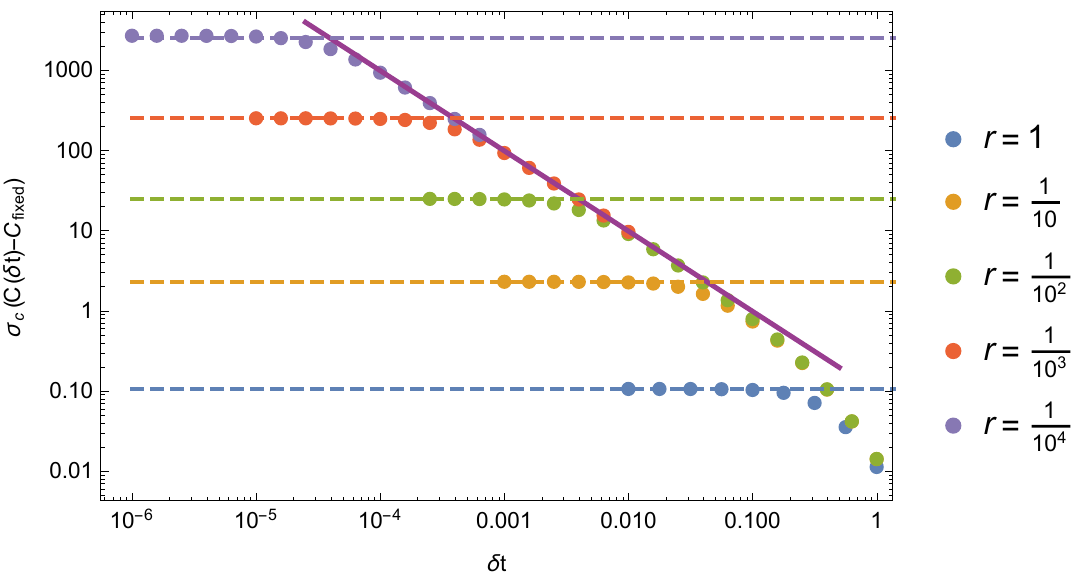}
\caption{(Colour online) Spatial correlator under a $d=5$ smooth quench at $t=0$ as a function of both $\dt$ and the distance separation $r$. In each case, we are subtracting the fixed mass correlator with $m^2=1/2$. The dashed lines correspond to computing the instantaneous quench correlator at $t=0$ for the different separations $r$, that is the same as computing the fixed mass correlator with $m=m_{in}=1$. The purple solid line shows the analytic leading order contribution to $<{\phi^2}>$, given by Table (\ref{fastresults}). This figure is taken from \cite{dgm3}.}
\label{early-time5}
\end{figure}

\begin{figure}[!h]
\centering\includegraphics[width=4.5in]{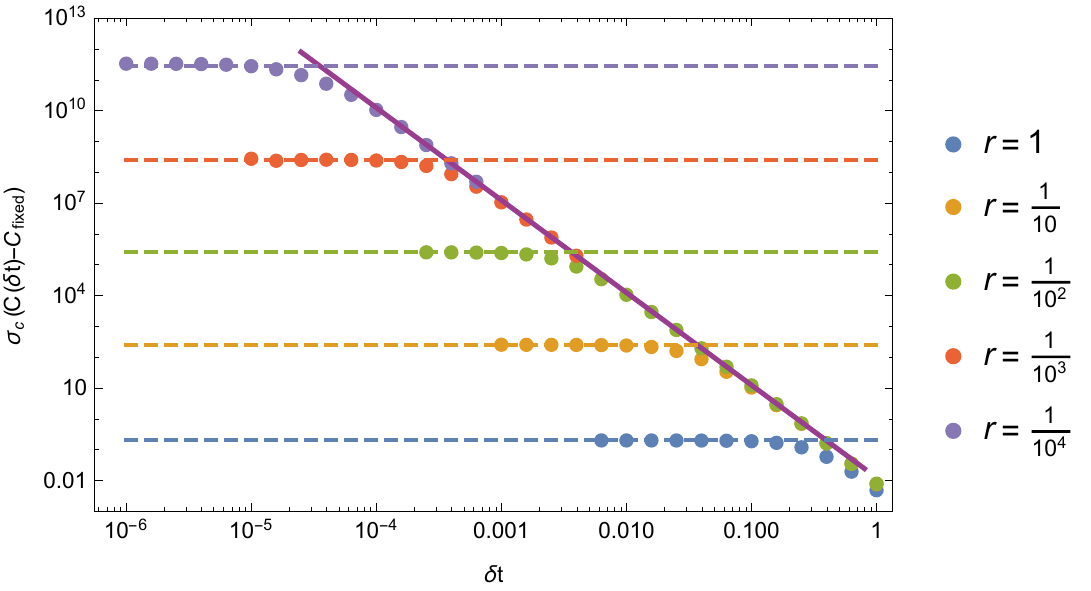}
\caption{(Colour online) Same as in Figure (\ref{early-time5}) for a $d=7$ smooth quench. This figure is taken from \cite{dgm3}.}
\label{early-time7}
\end{figure}

The most interesting behavior, however, happens for early times.  Figure (\ref{early-time5}) shows the result of the $d=5$ correlator at $t=0$ with a fixed mass contribution subtracted, as a function of the quench time $\dt$. The different colors are for different spatial separations. What we find is the following. For $r / \dt \gg 1$ the correlator staurates as a function of $\dt$. However as $\dt$ becomes of order $r$, the magnitude of the correlator scales as a function of $\dt$ : $C(r,t) \sim \dt^{d-2\Delta}$ for any fixed $r$. This is precisely the scaling of the composite operator $\phi^2$ discussed in the previous sections. For even larger $\dt$ the behavior departs from this fast quench scaling. The results are similar for finite $t$ which is close enough to the time when the quench ends. Figure (\ref{early-time7}) shows the result for $d=7$.

This scaling of correlation functions clearly shows that there is an intemediate regime between instantaneous quench and slow quench where a novel universal scaling behavior holds.

\subsubsection{Excess Energy}

Another UV finite object of physical interest is the excess energy density produced by the quench. This is defined by the quantity
\ben
\Delta \cE = \cE (t \rightarrow\infty) - \cE_{ground} (t=\infty)
\label{7-1}
\een
where $\cE (t)$ denotes the energy density at any time $t$, while $\cE_{ground}$ is the ground state energy of the final theory. The latter is a well defined object since the final theory has a constant coupling.

For any general theory we expect this quantity to be UV finite. The reason is the following. As we have seen above the {\em renormalized} energy at any time can be obtained by taking the bare energy density and subtracting counterterms which are given the adiabatic expansion to the appropriate order.  The ground state energy of the final theory is the lowest order adiabatic answer. The counterterms differ from this lowest order answer by terms which contain time derivatives of the coupling. Therefore if there is a quench protocol where the coupling approaches a constant fast enough, these additional terms vanish as $t \rightarrow \infty$ - so that the ground state energy of the final theory in fact coincides with the late time counterterm. 

Let us examine this explicitly for mass quenches in free scalar field theory. It is easy to see that the expression for $\Delta \cE$ is given by 
\ben
\Delta \cE = \int \frac{d^{d-1}k}{(2\pi)^{d-1}}\omega_{out} |\beta_\vk|^2 = \frac{\Omega_{d-2}}{(2\pi)^{d-1}} \int_0^\infty dk~k^{d-2}~ \omega_{out} \frac{\sinh^2 (\pi \omega_- \dt)}{\sinh (\pi \omega_{in} \dt) \sinh (\pi \omega_{out} \dt)}
\label{7-3}
\een
where $\beta_\vk$ is the Bogoliubov coefficient which is explicitly given in (\ref{6-3}). It can be checked that for any finite $\dt$ the expression in (\ref{7-3}) is finite both in the IR and the UV. 

Consider a quench from a massive theory to a massless theory. We wish to compare (\ref{7-3})  with the expression for the excess energy for an instantaneous quench, which can be obtained by using the appropriate Bogoliubov coefficients in (\ref{6-4}),
\ben
\Delta \cE^{instant} =  \frac{\Omega_{d-2}}{(2\pi)^{d-1}} \int dk~k^{d-1}
\frac{(\sqrt{k^2+m^2}-k)^2}{4k\sqrt{k^2+m^2}} \,.
\label{7-4}
\een
The table (\ref{excess-energy}) gives the result for the two quantities (\ref{7-3}) and (\ref{7-4}) for $m\dt \ll1$ in different dimensions
\begin{table}[!h]
\caption{Comparison of Excess Energies}
\label{excess-energy}
\centering
\begin{tabular}{|c||c||c|}
\hline
 $d=2 $ & $\delta \cE^{\dt \rightarrow 0} = \frac{m^2}{16\pi}+ c_1 m^4 (\dt)^2 + \cdots $ & $\delta \cE^{instant} = \frac{m^2}{16\pi}$ \\ \hline
$d=3$ & $\delta \cE^{\dt \rightarrow 0} = \frac{m^3}{24\pi} + c_2 m^4 \dt + \cdots $ & $ \delta \cE^{instant} = \frac{m^3}{24\pi} $ \\ \hline
$d=4$ & $\delta \cE^{\dt \rightarrow 0} = c_3 m^4 \log (m\dt) +O(\dt) $ & $ \delta \cE^{instant} = \infty$ \\ \hline
$ d\geq 5 $ & $ \delta \cE^{\dt \rightarrow 0} = c_4 m^4 \dt^{4-d} +\cdots $ & $ \delta \cE^{instant} = \infty$ \\ \hline
\end{tabular}
\end{table}
Several features of the results stand out. First, the instantaneous quench excess energy is finite for $d=2,3$. As expected the $\dt \rightarrow 0$ limit of the smooth quench answer agrees with this finte quantity in these cases. The subleading term is consistent with the scaling relation 
\ben
\Delta \cE - \Delta \cE^{instant} \sim (\dt)^{4-d} \sim (\dt)^{d-2\Delta}
\label{7-5}
\een
For $d \geq 4$ however the instantaneous excess energy diverges. Now it is the leading contribution which obeys a scaling relation.

While these results are for free fields, we expect that the lesson is quite general. For a general quench of the form described in section 4.1 we expect that if $2\Delta \leq d$ the $\dt \rightarrow 0$ limit of a smooth quench yields a finite result for the excess energy (and agrees with that of an infinitesimal quench), and the leading correction obeys a universal scaling law. On the other hand, for $2\Delta > d$ the $\dt \rightarrow 0$ is divergent in a universal fashion, while the corresponding instantaneous quench result is UV divergent.

\section{Conclusion}
A combination of holographic and standard field theoretic techniques have recently led to valuable insight into universal scaling properties in quantum quench. While we have discussed continuum field theories in this contribution, we have recently found that exactly these different scaling behaviors appear in a class of solvable lattice models. For a fixed lattice spacing, Kibble Zurek scaling appears for quenches which are slow compared to physical mass scales, the fast smooth scaling appears for quenches which are fast compared to physical mass scales but slow compared to the scale of the inverse lattice spacing, while quenches which are at the scale of the cutoff behave like instantaneous quenches. Since cold atom experiments can now probe lattice systems directly, these different regimes could be experimentally observable.

\section*{Acknowledgment} I am grateful to the organizers of the Nambu Symposium for inviting me to speak at this stimulating conference. This contribution is based on my work with Pallab Basu, Diptarka Das, Damian Galante, Takeshi Morita, Robert Myers, Tatsuma Nishioka and Krishnendu Sengupta. I thank them for a very fruitful set of collaborations. I would also like to thank Hong Liu, Gautam Mandal, Al Shapere, Steve Shenker, Eva Silverstein, Shibaji Sondhi, Lenny Susskind and Tadashi Takayanagi for discussions. I have benfitted from conversations with the participants of NORDITA workshop on Holography and Dualities 2016, and Yukawa Institute for Theoretical Physics workshop on Quantum Information in String Theory and Many Body Systems where parts of the work were presented. I would like to thank the organizers of these workshops for hospitality. This work was supported in part by National Science Foundation grants  NSF-PHY-1214341 and NSF-PHY-1521045.


%

\end{document}